\documentclass[10pt, pre, aps, twocolumn, showpacs, nofootinbib]{revtex4-1}
\usepackage{url,ulem}
\usepackage{amsmath,amssymb}
\usepackage[dvipdfmx]{graphicx, color}
\usepackage{dcolumn}
\usepackage{bm}
\usepackage{xcolor}
\usepackage{ulem}

\newcommand{\ave}[1]{\left\langle #1 \right\rangle}

\begin{document}

\title{Reply to: Comment on: Discontinuous codimension-two bifurcation in a Vlasov equation}

\author{Yoshiyuki Y. Yamaguchi$^{1}$}
\email{yyama@amp.i.kyoto-u.ac.jp}
\author{Julien Barr{\'e}$^{2}$}
\email{julien.barre@univ-orleans.fr}
\affiliation{
  $^{1}$Graduate School of Informatics, Kyoto University, Kyoto 606-8501, Japan\\
  $^{2}$Institut Denis Poisson, Universit{\'e} d'Orl{\'e}ans, Universit{\'e} de Tours and CNRS, 45067 Orl{\'e}ans, France}
\maketitle

The Comment \cite{TPL} criticizes the bifurcation analysis performed in \cite{YB} on a Vlasov equation.
This criticism can be traced back to a discrepancy in the definition of the "paramagnetic phase".
Apart from this discrepancy, there is no conflict between \cite{TPL} and \cite{YB}.

\begin{widetext}
\begin{figure}[bt]
  \centering
   \includegraphics[width=18cm]{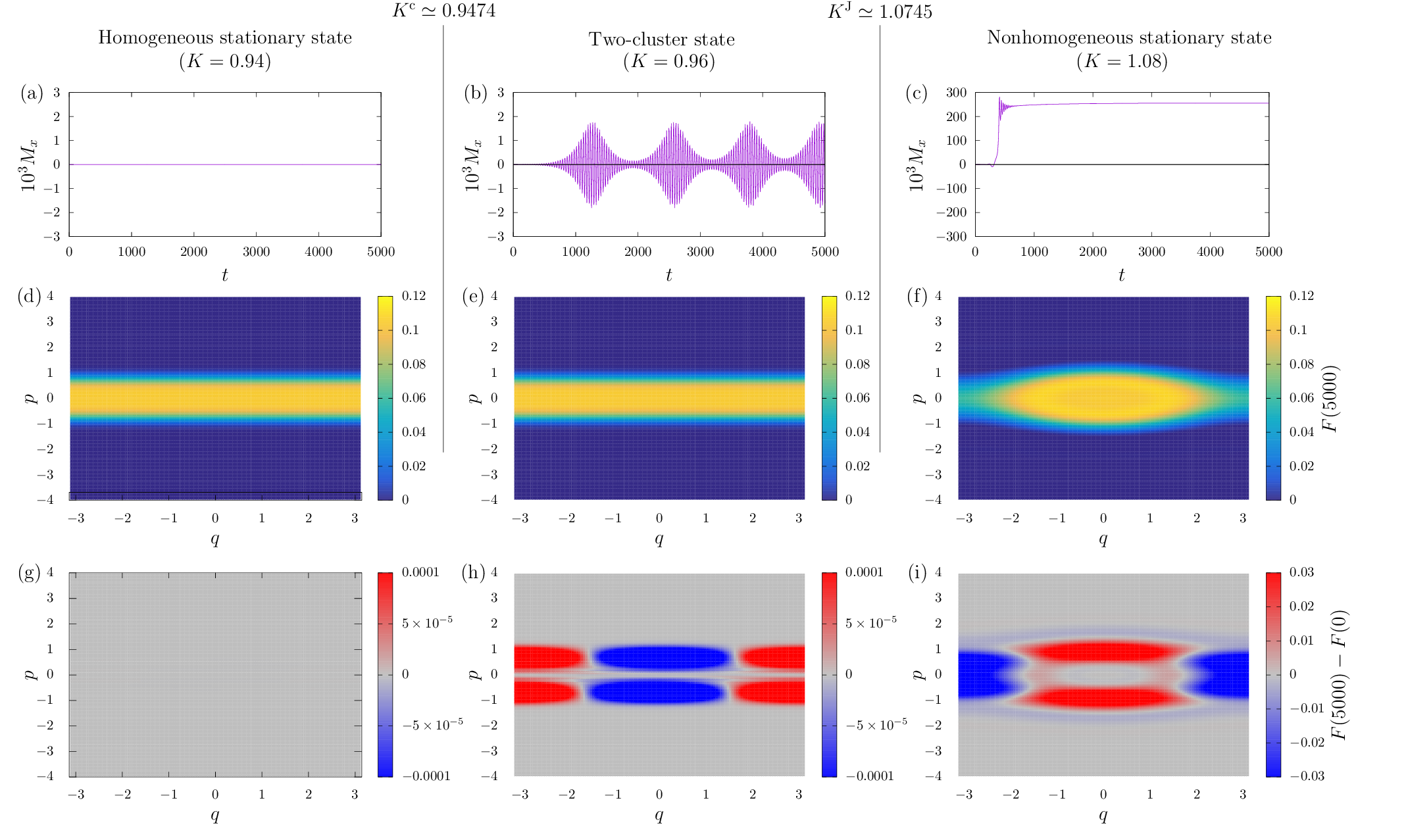}
   \caption{
     Two bifurcations in the generalized HMF model.
     The coupling constant $K_{1}$ (simply denoted by $K$) is the bifurcation parameter,
     and $K_{2}=0.5$.
     The two bifurcations are located at $K=K^{\rm c}$ and $K=K^{\rm J}$
     and separate the homogeneous stationary state, the two-cluster state,
     and the nonhomogeneous stationary state.
    $\beta_{2}=-0.3, \beta_{4}=3$ and $\epsilon=10^{-6}$ for the initial
    perturbed distribution \eqref{eq:Fepsilon}.
    (a,b,c): Temporal evolution of $10^{3}M_{x}(t)$.
    (d,e,f): Heat maps of $F(q,p,5000)$.
    (g,h,i): Heat maps of $F(q,p,5000)-F(q,p,0)$.
    The bifurcation parameter is $K=0.94$ (a,d,g),
    $K=0.96$ (b,e,h), and $K=1.08$ (c,f,i).
    Note that two clusters should be located around $|p|\simeq 0.174$
    [see Fig.\ref{fig:figure2}],
    which is the imaginary part of the most unstable eigenvalues
    and which corresponds to the shorter period in the panel (b).
    Mesh size is $128\times 128$ on a truncated phase space $(q,p)$
    which is $[-\pi,\pi]\times [-4,4]$.
  }
  \label{fig:figure}
\end{figure}
\end{widetext}

\vspace*{1em}
(1) The definition of the "paramagnetic phase" in \cite{TPL} is $\ave{M_{x}}=0$ (Def-1)
under the assumption of $M_{y}=0$ by symmetry; $\ave{M_{x}}$ denotes the time average of $M_{x}(t)=\iint \cos(q) F(q,p,t) dq\,dp$,
with $F(q,p,t)$ the time dependent position-velocity distribution function, governed by the Vlasov equation\footnote{For consistency with \cite{YB} we use $(q,p)$ as phase-space coordinates, and denote the magnetization by $(M_{x},M_{y})$ and $M$.}. 
However, the definition in \cite{YB} is that the distribution function $F$
 is spatially homogeneous (Def-2), namely $F$ does not depend on $q$.
A spatially homogeneous distribution (Def-2) implies $\ave{M_{x}}=0$ (Def-1),
but the converse is not true.
Hence Def-2 provides a finer classification of the states.
Indeed, if two clusters move in opposite directions [as in Fig.\ref{fig:figure}(h)],
$M_{x}(t)$ oscillates around zero [see Fig.\ref{fig:figure}(b)]
and Def-1 is satisfied while Def-2 is not.
This is what happens in Fig.7(c) of \cite{YB} in the range $K^{\rm c}<K<K^{\rm J}$
($K^{\rm c}$: critical point, $K^{\rm J}$: jump point),
where the order parameter $M$ is defined by $M=||(M_{x},M_{y})||$.
Increasing $K$, the two clusters merge when $K$ reaches $K^{\rm J}$
and the system is in a nonhomogeneous stationary state when $K>K^{\rm J}$.
This scenario is supported by Fig.\ref{fig:figure}(c,f) and
by Fig.11 of \cite{YB}.

\vspace*{1em}
(2) Based on Def-2, we find the two-cluster state in the range $K^{\rm c}<K<K^{\rm J}$.
  This third state is missed by adopting Def-1,
  while the existence of this intermediate oscillatory state is completely consistent with Figs. 2 and 4 of \cite{TPL}.
  The authors of \cite{TPL} insist that we interpreted the two-cluster oscillatory state
  as indicative of a continuous transition to the ferromagnetic phase.
  This statement is not true.
  What we have stated in \cite{YB} is a continuous bifurcation between the homogeneous
  stationary state and the two-cluster oscillatory state,
  where the bifurcation has been detected using $M$ instead of $M_{x}$.

\vspace*{1em}
(3) Dynamics in the two-cluster state is illustrated on Fig. \ref{fig:figure}(b),
and is typical of an oscillatory bifurcation:
$M_{x}(t)$ oscillates at a frequency given by the imaginary part of the unstable eigenvalue,
and with a slower varying amplitude.
Contrary to the claim in \cite{TPL}, the scaling law for this amplitude,
which represents the size of the clusters [see Fig.\ref{fig:figure}(h)],
is a crucial ingredient to understand this state.
Furthermore, we note that this two-cluster state persists for much longer computation times
than shown on Fig.\ref{fig:figure}.
These findings on the two-cluster state are actually not new:
two cluster solutions are constructed in \cite{BuchananDorning,YYY},
and the amplitude scaling is analyzed in \cite{Crawford,BMT}.

\begin{figure}
  \centering
  \includegraphics[width=8cm]{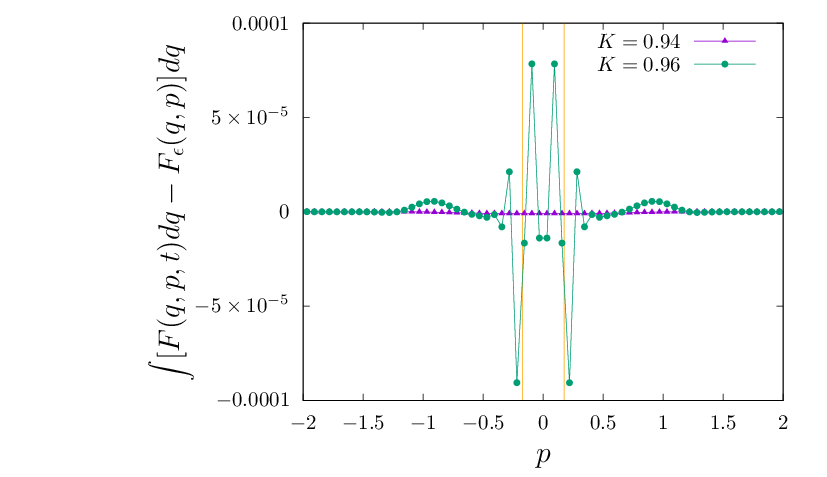}
  \caption{Modification of marginal distribution at time $t=5000$
    for $K=0.94$ (purple triangles) and $K=0.96$ (green circles).
    The two orange vertical lines mark $|p|=0.174$
    corresponding to the imaginary part of the most unstable eigenvalues at $K=0.96$.
    The parameters are the same as Fig.~\ref{fig:figure}.
  }
  \label{fig:figure2}
\end{figure}

All the above points are illustrated by Fig.\ref{fig:figure}, which expands Figs. 7 and 8 of \cite{YB}. The initial homogeneous distribution with perturbation is
\begin{equation}
  F_{\epsilon}(q,p) = C e^{-\beta_{2}p^{2}/2 - (\beta_{4}p^{2}/2)^{2}} [ 1 + \epsilon \cos(q) ],
  \label{eq:Fepsilon}
\end{equation}
where $C$ is the normalization factor to satisfy $\iint F_{\epsilon}(q,p) dqdp=1$,
We used the same parameters as in Fig.7(c) of \cite{YB}: $\beta_{2}=-0.3$, $\beta_{4}=3$,
and $\epsilon=10^{-6}$.
The truncated phase space $(q,p)\in (-\pi,\pi]\times [-4,4]$
is divided into an $L\times L$ mesh to perform a semi-Lagrangian algorithm,
which is a Vlasov solver \cite{deBuyl} and is used in \cite{YB},
where $L=128$ and the time step is $\Delta t=0.05$.
At $K=0.96$, the two clusters should be located around $|p|=0.174$,
which is the imaginary part of the most unstable eigenvalues.
This prediction is confirmed by the marginal distribution $\int F(q,p,t) dq$
at $t=5000$ as shown in Fig.\ref{fig:figure2},
while no bumps appear at $K=0.94$.

Summarizing, in the investigated Vlasov dynamics, there are three types of states:
a homogeneous stationary state, a nonhomogeneous stationary state,
and a two-cluster oscillatory state.
The last one can be captured by Def-2, adopted in \cite{YB},
but not by Def-1, adopted in \cite{TPL}.
The linear stability analysis identifies the continuous bifurcation point $K^{\rm c}$
between the homogeneous and two-cluster states.
Moreover, the nonlinear analysis in \cite{YB} approximately identifies
the discontinuous bifurcation point $K^{\rm J}$
between the two-cluster and the nonhomogeneous states.

We conclude that all Molecular Dynamics simulations in \cite{TPL} are consistent with \cite{YB},
when the third state (the two-cluster state) is considered.
The unique difference is the observation of multistability around the jump point
in the former due to finite-size fluctuations, which are absent in the Vlasov simulations of \cite{YB}.
There is no flaw in \cite{YB}.

\acknowledgements
Y.Y.Y. acknowledges support from JSPS KAKENHI Grant No. JP21K03402.


\end{document}